\documentclass[preprint]{JHEP} 
\usepackage{epsfig}
\usepackage{amsmath}
\usepackage{amssymb,amsfonts}
\newcommand{\Zint}{\mathbb{Z}}
\newcommand{\Real}{\mathbb{R}}
\newcommand{\be}{\begin{equation}}
\newcommand{\bea}{\begin{eqnarray}}
\newcommand{\eea}{\end{eqnarray}}
\newcommand{\beq}{\begin{equation}}
\newcommand{\ee}{\end{equation}}
\newcommand{\eeq}{\end{equation}}
\def\ba{\begin{eqnarray}}
\def\ea{\end{eqnarray}}
\def\K{{\cal K}}
\def\N{{\cal N}}
\def\O{{\cal O}}
\def\G{{\cal G}}

\def\tr{{\rm tr}}

\def\12{{1 \over 2}}
\def\32{{3\over 2}}
\def\52{{5\over 2}}
\def\72{{7\over 2}}
\def\92{{9\over 2}}
\def\AdS{{\rm AdS}}
\def\S{{\rm S}}
\def\g{g_{{\scriptscriptstyle Y \! M}}}
\def\gthree{g_{{\scriptscriptstyle Y \! M3}}}
\def\gfive{g_{{\scriptscriptstyle Y \! M5}}}
\def\A5{${\rm AdS}_5 \times {\rm S}^5$}
\def\B47{${\rm AdS}_4 \times {\rm S}^7$}
\def\C74{${\rm AdS}_7 \times {\rm S}^4$}
\newcommand{\irrep}[1]{\ensuremath{\boldsymbol{#1}}}

\title{Near-Extremal Correlators and Generalized
Consistent Truncation for AdS$_{4|7}\times$ S$^{7|4}$}

\author{Eric D'Hoker\\
Department of Physics\\
University of California, Los Angeles, CA 90095\\
E-mail: \email{dhoker@physics.ucla.edu} }

\author{Boris Pioline\footnote{On
leave of absence from LPTHE, Universit{\'e} Pierre et Marie Curie,
PARIS VI and Universit{\'e} Denis Diderot, PARIS VII, Bo\^{\i}te
126, Tour 16, 1$^{\it er}$ {\'e}tage, 4 place Jussieu, F-75252
Paris CEDEX 05, FRANCE}\\
Jefferson Physical Laboratory, Harvard University\\
Cambridge, MA 02138, USA\\
E-mail: \email{pioline@physics.harvard.edu}}

\preprint{\hepth{0006103}\\UCLA/2000/TEP/19\\HUTP-00/A24\\LPTHE-00-25}


\abstract{
We present conjectures for the space-time form and leading large $N$
dependence of extremal and near-extremal correlation functions in the
$\N=8$ superconformal Yang-Mills theory in $d=3$ as well as in the $(0,2)$
superconformal theory in $d=6$, using their gravity duals with M-theory
on $\AdS_4\times \S^7$ and $\AdS_7 \times \S^4$ respectively. As a key
part of the conjectures, we argue that the bulk couplings associated with
extremal and near-extremal field configurations in the
corresponding $\AdS_4$ and $\AdS_7$ gauged supergravities vanish. The
vanishing of these couplings constitutes a generalization of the property
of consistent truncation of the Kaluza-Klein modes.}

\begin{document}
\section{Introduction}

\medskip
\noindent {\it N=4 SYM and Type IIB supergravity on \A5}
\medskip

A number of remarkable conjectures on the factorization and coupling
constant dependence of correlators of local chiral operators have emerged
from the AdS/CFT conjecture \cite{Maldacena:1998re, Gubser:1998} between
Type IIB superstring theory on \A5 and $\N=4$ superconformal Yang-Mills
theory on
$\Real^4$. One important discovery is that certain correlators exhibit a
special factorized space-time form and are independent of the Yang-Mills
coupling, $\g$, a phenomenon usually referred to
as ``non-renormalization". Another surprising result is that certain
associated supergravity couplings in \A5 vanish, thereby extending the
usual property of consistent truncation. 
Most results obtained so far are on the correlators of 1/2 BPS
operators of $\N=4$ superconformal Yang-Mills theory, i.e. the theory
that corresponds to Type IIB superstring theory on \A5. The 
superconformal primary operators of this theory are denoted by
$\O_\Delta$ with dimension $\Delta$ and $SU(4)$ Dynkin label $(0,\Delta
,0)$. Normalizing the operators $\O_\Delta$ by their 2-point functions as
usual, the following conjectures have been proposed:


\newpage\noindent
{\bf Conjecture I : ${\cal N}$=4 SYM Correlators}
\begin{itemize}

\item[(1)] Non-renormalization of three-point functions $\langle
\O_{\Delta _1} (x_1) \O_{\Delta _2} (x_2) \O_{\Delta _3} (x_3)\rangle $
of single and multiple color trace 1/2 BPS operators $\O_{\Delta _i}$;

\item[(2)] Non-renormalization and factorization into a product of $n$
two-point functions of {\it extremal correlators} $\langle \O_{\Delta}
(x)  \O_{\Delta _1} (x_1) \cdots \O_{\Delta _n} (x_n) \rangle $ of single
and multiple trace 1/2 BPS operators $\O_{\Delta _i}$, whose
dimensions satisfy $\Delta = \Delta _1 +\cdots +\Delta _n$;

\item[(3)] Non-renormalization and decomposition into a sum of products of
a single non-renormalized three-point function and $n-1$ non-renormalized
two-point functions of {\it next-to-extremal correlators} $\langle
\O_{\Delta} (x) \O_{\Delta _1} (x_1) \cdots \O_{\Delta _n} (x_n) \rangle $
of single trace 1/2 BPS operators $\O_{\Delta _i}$, whose  dimensions
satisfy $\Delta +2 = \Delta _1 + \cdots + \Delta _n$;

\item[(4)] Decomposition into a sum of products of non-renormalized two-
and three-point functions and (in general, renormalized) higher point
functions of {\it sub-extremal correlators} $\langle \O_{\Delta} (x)
\O_{\Delta _1} (x_1) \cdots \O_{\Delta _n} (x_n) \rangle $ of single 1/2
BPS operators $\O_{\Delta _i}$ whose dimensions satisfy 
$\Delta +2m = \Delta _1 + \cdots + \Delta _n$ with $ 2\leq m \leq n-2$. 

\end{itemize}

The evidence for (1) derives from a detailed comparison between the 2- and
3-point functions calculated on the \AdS\ supergravity side, (using the
values for the supergravity couplings derived from the
supergravity field equations and action) and the free field values for the
same correlators on the superconformal Yang-Mills side \cite{Lee:1998bx,
Freedman:1999tz}. The \AdS\ supergravity expansion is valid for large $N$
and large 't~Hooft coupling
$\lambda = \g ^2 N$, so that equality between the correlators in the two
regimes supports the conjecture (1) at least in the large $N$
limit. Evidence for (1) for all values of $N$ is secured from the fact that
2- and 3-point functions suffer no corrections to first order
in $\g^2$ \cite{D'Hoker:1999tz, Skiba:1999im}.
Meanwhile, it was shown that 2-points also suffer no
corrections to order $\O(\g^4)$ \cite{Penati:1999ba}.  
Finally, methods of $\N=2$
analytic superspace \cite{Howe:1998zi}, combined with a recursive equation
of \cite{Intriligator:1999ff}, and
$\N=4$ superconformal invariance properties of the correlators provide
arguments that come close to being a general proof of
conjecture (1). (A loose end in the argument is related with the
existence of certain contact invariants, whose contribution has not been
ruled out so far.\footnote{We are grateful to Marc Grisaru for a
discussion on this point. See also
\cite{Penati:1999ba}.})

Part (2) of the conjecture was first proposed in \cite{D'Hoker:1999ea}
on the basis of evidence gathered from the \AdS\ supergravity side. A
priori, this evidence appears weaker than it was in the case of part (1),
since the explicit form of the \AdS\ gauged supergravity couplings has not
been evaluated directly from the supergravity action. Remarkably however,
there is an indirect argument that the supergravity couplings for extremal
arrangements of $n+1$-point couplings vanish as a consequence of the
finiteness of superstring theory on \A5 at the {\it string tree level},
as well as the fact that a Maldacena dual field theory exists at the
boundary of $\AdS_5$. Indeed, the
\AdS\ contact integrals appearing in the evaluation of the correlators
diverges precisely when the dimensions of the fields are extremal, so
that the associated supergravity couplings must vanish \cite{D'Hoker:1999ea}.
(See also \cite{Arutyunov:2000en} where this possibility was mentioned.) 
The extremal
4-point function has been shown to vanish by explicit calculation in
\cite{Arutyunov:1999fb}, thus confirming the proposed pattern. Meanwhile,
further evidence for (2) has accumulated from perturbative calculations
\cite{Bianchi:1999ie} and from $\N=2$ analytic superspace arguments
\cite{Eden:2000kw}.  

Part (3) was proposed in \cite{Eden:2000kw} based on $\N=2$ analytic
superspace argments, while further evidence from perturbation theory and
\AdS\ supergravity were given in \cite{Erdmenger:1999pz}. The 4-point 
next-to-extremal supergravity couplings were shown to vanish in 
\cite{Arutyunov:2000im}.
Part (4) was proposed in \cite{D'Hoker:2000dm} based on
perturbative and \AdS\ supergravity arguments, which furthermore lead to a
set of challenging conjectures on the structure of the \AdS\ gauged
supergravity theory:

\bigskip
\noindent
{\bf Conjecture II : AdS$_5 \times$S$^5$ Vanishing Near-Extremal
Couplings }
\begin{itemize}
\item  The bulk supergravity couplings $\G (\Delta, \Delta _1 ,\dots ,
\Delta _n)$ between the fields $s_{\Delta _i}$ dual to the 1/2 BPS single
trace operators  $\O_{\Delta _i}$ vanish for {\it near-extremal}
arrangements of dimensions $ \Delta +2m = \Delta _1 + \cdots + \Delta _n$
whenever $0 \leq m \leq n-2$,
\begin{equation}
\label{couplings}
\G (\Delta, \Delta _1 ,\dots , \Delta _n) =0  \, ;
\end{equation}
\end{itemize}
 
Note that this cancellation cannot merely be a property of $SU(2,2|4)$
superconformal invariance, since the superconformal Yang-Mills correlators
$\langle \O_{\Delta} (x) \O_{\Delta _1} (x_1) \dots$ $ \O_{\Delta _n} (x_n) 
\rangle$ in identically the same $SU(2,2|4)$ representation  
are non-vanishing.

There is further evidence for the validity of the \A5 supergravity
conjecture II from the well-known property of {\it consistent truncation}
of \AdS\ gauged supergravity. Consistent truncation means that Kaluza
Klein excitation modes of Type IIB on \A5 can systematically decouple
from the pure $\AdS_5$ gauged supergravity, since the latter is
a consistent theory by itself \cite{Gunaydin:1985qu}. This implies that
the supergravity couplings between $n$ pure \AdS\ supergravity fields and
any single Kaluza Klein excitation field must vanish, i.e. for all $n\geq
1$ we have
\begin{eqnarray}
\label{decpl}
\G (\Delta, 2_1, \dots ,2_n) =0\, .
\end{eqnarray}
For $\Delta > 2n$, $SU(4)_R$ group theory automatically guarantees the
vanishing of these couplings anyway, while, $SU(4)_R$ quadrality requires
$\Delta $ even. Thus, the non-trivial conditions imposed by consistent
truncation are the vanishing of the supergravity couplings for $\Delta = 4,
6, \dots , 2n$, which is precisely the contents of the conjecture
(\ref{couplings}) for the dimensions $\Delta _i =2$.

Though consistent truncation is by now a well-established property of the
supergravity field equations, a deeper understanding of its geometrical
significance still appears desirable. The \A5 supergravity conjecture II
(\ref{couplings}) provides a challenging generalization of standard
consistent truncation. As the vanishing of supergravity couplings in
(\ref{couplings}) is not merely a consequence of $SU(2,2|4)$ group theory,
its existence may point the way to additional hidden symmetries of gauged
supergravity.

\medskip
\bigskip
\noindent {\it $d=3,6$ SCFT 
and  11-dimensional SUGRA on $AdS_{4|7}\times S^{7|4}$}
\bigskip

In the present paper, we investigate to what extent factorization,
non-renormali-zation and generalized consistent truncation conjectures hold
for the AdS/CFT correspondences for M-theory on \B47 and \C74. The
11-dimensional supergravity on these spaces was considered long ago
in \cite{Biran:1984iy, Casher:1984ym, Pilch:1984xy, Gunaydin:1986tc} and
\cite{Pilch:1984xy, Gunaydin:1985wc, deWit:1984vq} respectively. The
property of consistent truncation was investigated in
\cite{deWit:1984vq,Nastase:1999cb}.

The associated superconformal field theory duals, $\N=8$ superconformal
Yang-Mills in $d=3$ and $(0,2)$ superconformal ``gauge theory" in $d=6$
respectively, are considerably less understood than their
$\N=4$, $d=4$ counterpart \cite{Aharony:1998rm}.
The key difficulty is the absence of
a freely adjustable coupling constant or marginal deformation. Instead,
both theories emerge as isolated strong coupling fixed points, with only
the size $N$ of the gauge group $SU(N)$ as a free parameter. Moreover, 
finite temperature computations, absorption cross-sections 
and anomalies reveal that the
effective number of degrees of freedom in the large $N$ limit behaves as
$N^{3/2}$ for $d=3$ and as $N^3$ for $d=6$, both radically different from
the customary field theory results \cite{Klebanov:1996un}.
On the \AdS\ side, the absence of marginal deformations implies that the
$1/N$ expansion will govern both the quantum corrections and the low
energy expansion of M-theory on \B47 or on \C74. 

The absence of any marginal deformations to the $d=3$ and $d=6$ theories
renders the issue of ``non-renormalization" of AdS/CFT correlators moot, as
there simply is no free coupling constant to be considered. However,
keeping in mind the analogy with the \A5 case, the questions of
``factorization" as well as of the ``vanishing of near-extremal
supergravity couplings" continue to be challenging.  It is these questions
that we shall investigate here.  In particular, we shall present evidence
for the following conjecture.

\bigskip
\noindent
{\bf Conjecture III : AdS$_{4|7} \times$S$^{7|4}$ Vanishing
Near-Extremal Couplings}
\begin{itemize}
\item  The bulk supergravity couplings $\G (\Delta, \Delta _1 ,\dots ,
\Delta _n)$ between the fields
$s_{\Delta _i}$ dual to the 1/2 BPS single trace operators $\O_{\Delta _i}$
vanish for {\it near-extremal} arrangements of 
dimensions $ \Delta + 2m {\cal K} =
\Delta _1 + \cdots + \Delta _n$ whenever $0 \leq m \leq n-2$,
\begin{equation}
\label{couplings2}
\G (\Delta, \Delta _1 ,\dots , \Delta _n) =0  \, ;
\end{equation}
\end{itemize}

Here $\K$ is the unit of dimension for superconformal primaries,
$\K=1/2$ for AdS$_4$ and $\K=2$ for AdS$_7$.
Just as in the case of Type IIB supergravity on \A5, this 
conjecture encompasses the vanishing relations that are equivalent 
to consistent truncation, and generalizes consistent truncation 
in a non-trivial way. 

Our main argument for this conjecture will come from the finiteness
of the extremal correlator of boundary superconformal primaries: on
the AdS gravity side, this correlator arises from a number 
of tree-level exchange diagrams plus one contact diagram.
All exchange diagrams are finite, but the integral on
the position of the vertex in the contact diagram diverges when 
the vertex approaches the boundary insertion with highest conformal
dimension. Hence the vertex should vanish for the correlator
to be finite. While this argument will be spelled out in detail
for the \B47 and \C74 cases in Section 4, it should be noted that it
is expected to apply more generally for any weakly coupled supergravity
on a product space $\AdS_{d+1} \times {\cal M}$, provided this theory
admits a finite conformal quantum field theory dual.
Assuming that a discrete spectrum of operators $\O_\Delta$ exists, (as
expected for ${\cal M}$ compact) the supergravity coupling between
the dual fields $s_{\Delta _i}$ whose exact quantum dimensions  satisfy
$\Delta = \Delta _1 + \dots + \Delta _n$ will be forced to vanish by 
the same reasoning as above. Such linear
relations between dimensions are the rule in supersymmetric theories, but
it is conceivable that the argument  could be generalized to situations
with softly or spontaneously broken supersymmetry as well. It is  expected
to apply in  particular to Type IIB superstrings on
$\AdS_3 \times S^3 \times T^4$ with 32 supercharges, to $\AdS_3 \times
S^3 \times K_3$ with 16 supercharges and to $\AdS_5 \times S^5/\Gamma$ with
16 or 8 supercharges amongst other examples.

\medskip
\bigskip
\noindent {\it Outline}
\bigskip

The remainder of the paper is organized as follows. In Section 2, we review the
$d=3,6$ superconformal theories, including the structure of their 1/2 BPS
operators.  In Section 3, 
we present general convergence and divergence criteria
for tree level integrals on \AdS\ space-times. In Section 4, we use the
assumption of finiteness to show that extremal supergravity couplings must
vanish, and we derive a general factorized form of extremal correlation
functions 1/2 BPS operators.  In Section 5, we present conjectures on the
factorization and vanishing of near-extremal supergravity couplings.

\section{The $d=3, 6$ superconformal theories}

The $d=3$ and $d=6$ superconformal theories with 16 supercharges 
describe the world-volume dynamics of a stack of $N$ coincident M2-branes
or M5-branes embedded in flat eleven-dimensional Minkowski space. These 
interacting theories can be obtained by a renormalization flow from
the more familiar supersymmetric gauge theories describing the 
dynamics of non-coincident D2 and D4 branes at weak coupling respectively
(see \cite{Seiberg:1998ax} for a review). Indeed, the three-dimensional
$\N=8$ super-Yang-Mills theory living on the  coincident D2-branes
is strongly coupled at energies much smaller than $\gthree^2=g_s/l_s=
l_p^3/R_s^2$, where $l_p$ is the eleven-dimensional Planck length
and $R_s=g_s l_s$ the size of the eleven dimension,
and flows to an interacting infrared fixed point 
with $\N=8$ superconformal symmetry 
in the eleven-dimensional 
decompactification limit $R_s/l_p\to 0$ \cite{Banks:1997my}.
On the contrary, the five-dimensional $\N=4$ super-Yang-Mills theory
on the coincident D4-branes is Gaussian in the infrared, but strongly 
coupled at energies much bigger than $1/\gfive ^2=1/(g_s l_s)=1/R_{s}$.
At this scale, a new dynamically generated dimension opens up, whose
momentum excitations are the Yang-Mills instantons of 
mass $1/\gfive ^2$ \cite{Rozali:1997cb}.
The ultraviolet behaviour is controlled by a $d=6$ $(0,2)$ 
superconformal theory, which is also the world-volume theory of
the M5-brane. The same theory can also be obtained as the decoupling
limit $g_s\to 0, l_s\to 0$ of the type IIA 
NS5-brane \cite{Strominger:1996ac}, or as 
the type IIB string theory compactified on a $A_{N-1}$ 
singularity \cite{Witten:1995zh}.
We now briefly recall properties of these superconformal fixed points,
based on their symmetry algebras, which are two different
real forms of the orthosymplectic superalgebra $OSp(8|4)$,
whose representations can be analyzed as in \cite{Dobrev:1985vh}.

\subsection{The $d=3$ $N=8$ superconformal theory}

The $d=3$ theory has conformal group $SO(3,2)\sim Sp(4)$,
extended with 8 odd generators in the pseudo-real four-dimensional spinor
representation of $SO(3,2)$, rotated into each other\footnote{We take 
the convention  that the supersymmetry charges transform
as a spinor \irrep{8_s} of $SO(8)$. }
by an R-symmetry $SO(8)$.
In the $d=3$ SYM theory, only a $SO(7)$ subgroup of $SO(8)$ is realized
linearly, and the full $SO(8)$ symmetry emerges dynamically only in the
strong coupling limit. The theory contains one gauge field $A_\mu$, one
fermion $\lambda^a$ in the spinor representation \irrep{8} of $SO(7)$ and
seven scalars $X^I$, $I=1,\dots,7$, in the vector representation
\irrep{7}, all  taking  values in the Lie algebra of $SU(N)$. 
The microscopic Lagrangean is given by
\begin{equation}
{\cal L}=
\frac{1}{\gthree^2} \tr \left\{
F_{\mu\nu}^2 + (\nabla_\mu X^I)^2 + [X^I,X^J]^2
+ \lambda^a \gamma^\mu \nabla_\mu\lambda^a 
+ \lambda^a \Gamma^I_{ab} X^I \lambda^b
\right\}
\end{equation}
where $\Gamma^I$ and $\gamma_\mu$ are $SO(7)$ (internal) and $SO(2,1)$
(space-time) gamma matrices respectively.
In the free theory, the gauge field can be dualized into an eighth scalar
$X^8$, in terms of which  the free Lagrangean exhibits the full
$SO(8)$ R-symmetry. The field content is now 8 scalars $X^i$, $i=1,\dots
,8$, and eight fermions $\lambda _a $, $a=1,\dots, 8$. Denoting the
$SO(8)$ gamma matrices by $\Gamma ^i$, we have the following supersymmetry
transformations,
\begin{eqnarray}
\delta X^i= -i \epsilon^a \Gamma^i_{ab} \lambda^b\ ,\quad
\delta \lambda_a =  \Gamma^i_{ab} \gamma^{\mu} \partial_\mu X^i \epsilon^b
\end{eqnarray}

The complete spectrum at the infrared fixed point is not known, but the
chiral (or BPS) operators can be followed from weak coupling, since 
their dimensions are protected from quantum corrections. They form
infinite dimensional unitary representations of $OSp(2,6|4)$,
for which oscillator \cite{Gunaydin:1985wc} or harmonic superspace 
constructions \cite{Ferrara:2000eb} are available.
These representations are built by applying the supersymmetry generators
on a lowest-weight vector (or superconformal primary operator, SCPO).
This yields a finite number of chiral primary operators (CPO), each of
which heads an infinite tower of conformal descendents.
Here, we shall be interested in 1/2 BPS operators only, defined by
\begin{eqnarray}
\label{ok}
\O_k = \tr X^k \equiv {\rm Str} (X^{i_1} \cdots X^{i_k})
\end{eqnarray}
with dimension $\Delta=k/2$ protected from quantum corrections. 
Str stands for the symmetrized color
trace, and it is assumed that the indices $i_j$ are made traceless.
The fields $X^i$ denotes the free field with $SO(8)$ Dynkin label
$(k,0,0,0)$ \footnote{Hence
$\O_k$ carries a charge $(-)^k$ under the subgroup $\Zint_2$ in the
center $\Zint_2\times \Zint_2$ of $Spin(8)$.}. Its superconformal descendents
are listed in Table 1, and their dimensions are easily computed from the
free field dimensions $[X]=1/2, [\lambda]=1, [F]=3/2$. 
Non-symmetric or traceful representations do
not give rise to 1/2 BPS operators. On the other hand, multi-trace
operators in the same $(k,0,0,0)$ are also 1/2 BPS operators 
\cite{Andrianopoli:1999ut}, and mix
with the single trace ones only at subleading order in $1/N$.

\begin{table}[t]
\begin{center}
\begin{tabular}{|l|c|c|c|c|c|c|c|} \hline 
Free Field Operator   & desc & SUGRA$_4$ & 2$\times$ dim & spin$^P$  & 
$SO(8)$ &
 lowest reps
                \\ \hline \hline
$\O _k         \sim    X^k$, $k\geq 2$        
                  & -- 
                  & $h_{\alpha}^{\alpha},\ \ D_\alpha D_\beta  h^{\alpha
\beta}$ 
                  & $k$
                  & $0^+$    
                  & $(k,0,0,0)$
                  & {\bf 35}$_v$,{\bf 112}$_v$
                \\ \hline
$\O _k ^{(1)}    \sim  \lambda X^k$, $k\geq 1$
                  & $Q$  
                  & $\psi _{\alpha}$ 
                  & $k+2$
                  & $\12$ 
                  & $(k,0,1,0)$
                  & {\bf 56}$_s$,{\bf 224}$_{vc}$
                \\ \hline
$\O _k ^{(2)}    \sim  \lambda \lambda X^k$ 
                  & $Q^2$  
                  & $a_{\alpha \beta \gamma} $
                  & $k+4$
                  & $0^-$ 
                  & $(k,0,2,0)$
                  & {\bf 35$_c$},{\bf 224$_{cv}$}
                \\ \hline
$\O _k ^{(3)}   \sim   \lambda \lambda X^k$   
                  & $Q^2$
                  & $ h_{\mu \alpha}, \ \ a_{\mu \alpha \beta}$
                  & $k+4$
                  & $1^-$ 
                  & $(k,1,0,0)$
                  & {\bf 28},{\bf 160}$_{v}$
                \\ \hline
$\O _k ^{(4)} \!  \sim \!  \lambda \lambda \lambda X^k$
                  & $Q^3$ 
                  & $\psi _\alpha $
                  & $k+6$
                  & $\12$ 
                  & $(k,1,1,0)$
                  & {\bf 160$_c$},{\bf 840$_s$}
                \\ \hline
$\O _k ^{(5)}   \sim   \lambda \lambda \lambda X^k$ 
                  & $Q^3$
                  & $\psi _\mu$
                  & $k+6$
                  & $\32$ 
                  & $(k,0,0,1)$
                  & {\bf 8}$_s$,{\bf 56}$_c$
                \\ \hline
$\O _k ^{(6)}    \sim  (\partial X) ^2 X^k$   
                  & $Q^4$
                  & $h_{\mu \nu}$
                  & $k+6$
                  & $2^+$ 
                  & $(k,0,0,0)$
                  & {\bf 1},{\bf 8}$_v$, {\bf 35}$_v$
                \\ \hline
$\O _k ^{(7)}   \sim   (\partial X) \lambda \lambda X^k$ 
                  & $Q^4$
                  & $a _{\mu \alpha \beta}$
                  & $k+7$
                  & $1^+$ 
                  & $(k,0,1,1)$
                  & {\bf 56}$_v$,{\bf 350}
                \\ \hline
$\O _k ^{(8)}    \sim  \lambda \lambda \lambda \lambda X^k$
                  & $Q^4$
                  & $h_{(\alpha \beta)}$
                  & $k+8$
                  & $0^+$ 
                  & $(k,2,0,0)$
                  & {\bf 300},{\bf 1400}
                \\ \hline
$\O _k ^{(9)}   \sim  (\partial X)^2 \lambda X^k$ 
                  & $Q^5$
                  & $\psi _\mu$
                  & $k+8$
                  & $\32$ 
                  & $(k,0,1,0)$
                  & {\bf 8}$_c$,{\bf 56}$_s$
                \\ \hline
$\O _k ^{(10)}   \sim  (\partial X) \lambda \lambda \lambda  X^k$ 
                  & $Q^5$
                  & $\psi _\alpha$
                  & $k+9$
                  & $\12$ 
                  & $(k,1,0,1)$
                  & {\bf 160}$_s$,{\bf 840}$_c$
                \\ \hline
$\O _k ^{(11)}   \sim  (\partial X)^2 \lambda \lambda X^k$ 
                  & $Q^6$
                  & $ h_{\mu \alpha}, \ \ a_{\mu \alpha \beta}$
                  & $k+10$
                  & $1^-$ 
                  & $(k,1,0,0)$
                  & {\bf 28},{\bf 160}$_v$
                \\ \hline
$\O _k ^{(12)}   \sim  (\partial X)^2 \lambda \lambda X^k$ 
                  & $Q^6$
                  & $a_{\alpha \beta \gamma}$
                  & $k+10$
                  & $0^-$ 
                  & $(k,0,0,2)$
                  & {\bf 35$_c$},{\bf 224$_{cv}$}
                \\ \hline
$\O _k ^{(13)}   \sim  (\partial X)^3 \lambda X^k$ 
                  & $Q^7$
                  & $\psi _\alpha$
                  & $k+11$
                  & $\12$ 
                   & $(k,0,0,1)$
                  & {\bf 8}$_s$,{\bf 56}$_c$
                \\ \hline
$\O _k ^{(14)}   \sim  (\partial X)^4 X^k$ 
                  & $Q^8$
                  & $h_{\alpha}^{\alpha},\ \ D_\alpha D_\beta  h^{\alpha
\beta}$
                  & $k+12$
                  & $0^+$ 
                  & $(k,0,0,0)$
                  & {\bf 1},{\bf 8}$_v$, {\bf 35}$_v$
                \\ \hline
\end{tabular}
\end{center}
\caption{$\AdS_4 \times {\rm S}^7$ Supergravity fields and $SO(3,2) \times
SO(8)$ quantum numbers.  The range of $k$ is $k\geq 0$, unless otherwise 
specified. }
\label{table:1}
\end{table}

\subsection{The $d=6$ (0,2) superconformal theory}

The situation with the $d=6$ $(0,2)$ superconformal theory 
is very similar. The conformal group $SO(6,2)$ is extended with
4 odd generators transforming as symplectic-Weyl-Majorana spinors
of $SO(6,2)$, rotated into each other by an R-symmetry $USp(4)\sim SO(5)$.
In contrast to the $d=3$ case, the D4-brane theory exhibits 
the full R-symmetry, since it contains one gauge field $A_\mu$, five scalars
$X^I$ in the \irrep{5} of $USp(4)$ and four pseudoreal 
fermions $\lambda$ in the \irrep{4}
of $USp(4)$.
It is however more convenient to describe the spectrum in terms
of the free $d=6$ $(0,2)$ tensor multiplet, which makes the
six-dimensional Lorentz symmetry manifest. This theory contains one
two-form with  self-dual field strength $H^+_{\mu\nu\rho}$, 
which reduces to the $d=5$
field strength $F_{\mu\nu}=H_{\mu\nu 5}$, four symplectic Majorana-Weyl
fermions in the \irrep{4} of $USp(4)$, and five scalars. The
supersymmetry transformations in the free theory are given  by
\cite{Howe:1983fr}
\begin{eqnarray}
\delta X^I&=& - \epsilon^a \Gamma^I_{ab} \lambda^b\ ,\quad
\delta H_{\mu\nu\rho}^+= \epsilon_a \Gamma_{[\mu\nu}^{ab} 
\partial_{\rho]} \lambda_b\ ,\quad \nonumber\\
\delta \lambda_a &=&   \frac14 \gamma^{\mu} \partial_\mu X^I
\Gamma^I_{ab} \epsilon^b
-\frac{1}{12} H_{\mu\nu\rho}^+\Gamma^{\mu\nu\rho}_{ab}\epsilon^b
\end{eqnarray}
All fields take values in the Lie algebra of $SU(N)$,
although it is as yet unknown how to consistently switch on
interactions. Still it is possible to follow the spectrum of BPS
operators from zero-coupling to the superconformal ultraviolet 
fixed point. The 1/2 BPS chiral operators are obtained
from the superconformal primary operator
\be
\O_k=\tr X^k \equiv {\rm Str} (X^{i_1} \cdots X^{i_k})
\end{equation}
now in the symplectic-traceless completely symmetric tensor product
of $k$ fundamentals of $USp(4)$, with Dynkin labels $(0,k)$ and dimension
$2k$. In particular, $\O_k$ carries a charge $(-)^k$ under the center of
$USp(4)$. The dimensions of the descendents are computed using the
free-field dimensions $[X]=2, [\lambda]=5/2$ and $[H_+]=3$.

\begin{table}[b]
\begin{center}
\begin{tabular}{|l|c|c|c|c|c|c|c|} \hline 
Free field Operator   & desc & SUGRA$_7$ & dim & $SU(4)^*$  & $USp(4)$ &
 lowest reps
                \\ \hline \hline
$\O _k         \sim    X^k$, $k\geq 2$        
                  & -- 
                  & $h_{\alpha}{}^{\alpha}$ 
                  & $2k$
                  & $(0,0,0)$    
                  & $(0,k)$
                  & {\bf 5},{\bf 14},{\bf 30}
                \\ \hline
$\O _k ^{(1)}    \sim  \lambda X^k$, $k\geq 1$
                  & $Q$  
                  & $\psi _{\alpha}$ 
                  & $2k+\52$
                  & $(1,0,0)$ 
                  & $(1,k)$
                  & {\bf 4},{\bf 16},{\bf 40}
                \\ \hline
$\O _k ^{(2)}    \sim  H_+ X^k$ 
                  & $Q^2$  
                  & $A_{\mu \nu \rho} $  
                  & $2k+3$
                  & $(2,0,0)$ 
                  & $(0,k)$
                  & {\bf 1},{\bf 5},{\bf 14}
                \\ \hline
$\O _k ^{(3)}   \sim   \lambda \lambda X^k$   
                  & $Q^2$
                  & $ h_{\mu \alpha} \ \ B_{\mu \alpha}$
                  & $2k+5$
                  & $(0,1,0)$ 
                  & $(2,k)$
                  & {\bf 10},{\bf 35},{\bf 81}
                \\ \hline
$\O _k ^{(4)}   \sim  H_+ \lambda   X^k$ 
                  & $Q^3$
                  & $\psi _{\alpha}$
                  & $2k+{11 \over 2}$
                  & $(1,0,0)$ 
                  & $(1,k)$
                  & {\bf 4},{\bf 16},{\bf 40}
                \\ \hline
$\O _k ^{(5)}    \sim  \lambda \lambda \lambda X^k$ 
                  & $Q^3$
                  & $\psi _\mu$
                  & $2k+{15\over 2}$
                  & $(1,1,0)$ 
                  & $(3, k)$
                  &  {\bf 20},{\bf 64}, {\bf 140}
                \\ \hline
$\O _k ^{(6)}    \sim  H_+^2 X^k$   
                  & $Q^4$
                  & $h_{\mu \nu}$
                  & $2k+6$
                  & $(0,2,0)$ 
                  & $(0,k)$
                  & {\bf 1},{\bf 5},{\bf 14}
                \\ \hline
$\O _k ^{(7)}   \sim  H_+ \lambda \lambda X^k$ 
                  & $Q^4$
                  & $A_{\mu \alpha \beta} \ \ B_{\mu \alpha}$
                  & $2k+8$
                  & $(1,0,1)$ 
                  & $(2,k)$
                  & {\bf 10},{\bf 35},{\bf 81}
                \\ \hline
$\O _k ^{(8)}    \sim  \lambda \lambda \lambda \lambda X^k$   
                  & $Q^4$
                  & $h_{(\alpha \beta)}$ 
                  & $2k+10$
                  & $(0,0,0)$ 
                  & $(4,k)$
                  & {\bf 35'},{\bf 105}
                \\ \hline
$\O _k ^{(9)}   \sim  H_+^2  \lambda X^k$ 
                  & $Q^5$
                  & $\psi _\mu$
                  & $2k+{17\over 2}$
                  & $(1,1,0)$ 
                  & $(1,k)$
                  & {\bf 4},{\bf 16},{\bf 40}
                \\ \hline
$\O _k ^{(10)}   \sim  H_+ \lambda  \lambda \lambda X^k$ 
                  & $Q^5$
                  & $\psi _{\alpha}$
                  & $2k+{21\over 2}$
                  & $(1,0,0)$ 
                  & $(3,k)$
                  & {\bf 20},{\bf 64}, {\bf 140}
                \\ \hline
$\O _k ^{(11)}   \sim  H_+^3 X^k$ 
                  & $Q^6$
                  & $h_{\mu \alpha} \ \  B_{\mu \alpha}$
                  & $2k+9$
                  & $(0,1,0)$ 
                  & $(0,k)$
                  & {\bf 1},{\bf 5},{\bf 14}
                \\ \hline
$\O _k ^{(12)}   \sim  H_+^2 \lambda \lambda  X^k$ 
                  & $Q^6$
                  & $A_{\mu \nu \rho}$
                  & $2k+11$
                  & $(1,0,1)$ 
                  & $(2,k)$
                  &   {\bf 10},{\bf 35},{\bf 81}
                \\ \hline
$\O _k ^{(13)} \!  \sim   H_+ ^3 \lambda  X^k$ 
                  & $Q^7$
                  & $\psi _\alpha$
                  & $2k+{23\over 2}$
                  & $(1,0,0)$ 
                  & $(1,k)$
                  & {\bf 4},{\bf 16},{\bf 64}
                \\ \hline
$\O _k ^{(14)}   \! \sim  H_+ ^4 X^k$ 
                  & $Q^8$
                  & $h_\alpha {}^\alpha$
                  & $2k+12$
                  & $(0,0,0)$ 
                  & $(0,k)$
                  & {\bf 1},{\bf 5},{\bf 14}
                \\ \hline
\end{tabular}
\end{center}
\caption{$\AdS_7 \times {\rm S}^4$ Supergravity fields and 
$SO(6,2) \times USp(4)$ quantum numbers. 
The range of $k$ is $k\geq 0$, unless otherwise specified. }
\label{table:2}
\end{table}

\subsection{AdS dual and consistent truncation}

The BPS spectrum is the only information about the superconformal
fixed point that can be reliably computed using the free field
theories. On the other hand, Maldacena's conjecture gives a
dual description of the large $N$ limit of these $d=3$ and $d=6$ fixed 
points, in terms of eleven-dimensional supergravity in the near-horizon
geometry of the M2 and M5-brane, namely \B47
and \C74 respectively \cite{Maldacena:1998re}. 
This description is reliable when the radius of the
4-sphere $R_4=N^{1/6}l_p$ or 7-sphere $R_7=N^{1/3}l_p$ is much larger than 
the Planck length $l_p$, {\it i.e.} at large $N$. 
The spectrum of
chiral primary fields on the gauge theory is easily matched 
\cite{Aharony:1998rm} to the spectrum
of Kaluza-Klein states of 11d supergravity compactified on the sphere,
which was worked out in \cite{Biran:1984iy} and \cite{Pilch:1984xy} 
respectively, as shown
in Table 1 and 2. The operator on the gauge theory side is
represented by its free-field version, using the dualized
representation for $d=3$ and the self-dual tensor multiplet
for $d=6$, and a symmetrized traceless
trace in the adjoint representation is
understood. In both cases, 
the operator with $k=1$ is a pure gauge (doubleton) mode
on the supergravity side, which justifies looking at $SU(N)$ theories only.
The operator with $k=2$ and its descendents correspond to the $\N=8$
supergravity multiplet in the limit of flat Minkowski space, and include
the graviton, gravitini, gauge fields, fermions and scalars in the
$\irrep{1}\oplus \irrep{8_s} \oplus \irrep{28} \oplus \irrep{56_s} 
\oplus \irrep{35_v}\oplus \irrep{35_c}$ for the M2-brane,
and $\irrep{1}\oplus \irrep{4}\oplus \irrep{10} \oplus \irrep{16} 
\oplus \irrep{14}$ for the M5-brane.
Multitrace operators correspond to multiparticle states on the gravity
side.

In addition to yielding the spectrum at the superconformal point, the
AdS dual description also allows to extract correlation functions
between chiral primaries: in the large $N$ limit, they are given by
diagrams whereby the boundary fields propagates from the boundary to the 
bulk and interact locally as well as by exchange of bulk modes.
This computation requires identifying the gravity modes to which the
gauge theory operators couple, and their $n$-point couplings in the
bulk supergravity.
Following the \A5 computation of \cite{Lee:1998bx}, three-point
functions for superconformal primaries have been extracted 
in \cite{Corrado:1999pi}. This has been extended
to a class of superconformal descendents 
in \cite{Bastianelli:2000vm}.
In particular, these results show that extremal 3-point functions
vanish in the large $N$ approximation.
Three-point correlators of stress tensors have also been computed
using the AdS/CFT correspondence, and turn out to be given by
the free-field result \cite{Bastianelli:1999ab}.

The vanishing of extremal 3-point functions implies in particular
that the coupling between two massless states and a massive Kaluza-Klein
state vanishes. This fact raises the possibility of truncating
the spectrum in order to define a reduced gauge supergravity theory
on AdS$_4$ or AdS$_7$, as is customary in Kaluza-Klein reductions on tori.
Indeed, it has been shown by de Wit and Nicolai long ago in the 
\B47 case \cite{deWit:1984vq}, 
and more recently in \C74 \cite{Nastase:1999cb},
that one may truncate the Kaluza-Klein
spectrum to the massless modes in such a way that any solution
of the truncated theory can be lifted to a solution of 11d supergravity.
The vanishing of extremal 3-point couplings is only a necessary
condition for this to hold, and more generally all couplings
involving one massive mode and $n$ massless modes should vanish, as
in (\ref{decpl}). In this paper, we shall argue for an even more general 
decoupling occurring for near-extremal configurations, 
which hints to a deeper formulation
of consistent truncation.

\section{Convergence Criteria for AdS integrals}

As for the treatment of near-extremal correlation functions in the \A5
case \cite{D'Hoker:1999ea, D'Hoker:2000dm}, we shall also here make heavy
use of the divergence properties of $\AdS _{d+1}$ integrals, in
particular for $d=3,6$.  Thus, we extend the arguments of
\cite{D'Hoker:1999ea, D'Hoker:2000dm} to general $d$ in order to
emphasize that the divergence structure is independent of $d$.

Throughout, we analytically continue to Euclidean \AdS\, which may be
represented by the upper half space $\AdS_{d+1} = \{ (z_0, \vec{z}). \
z_0 >0, \ \vec{z} \in \Real^d \}$, with the Poincar\'e metric
$ds^2 =  ( dz_0^2 + d\vec{z}^2 )/z_0^2$.
All \AdS\ integrals of interest to us are associated with correlation 
functions of the superconformal primary operator of a 1/2 BPS multiplet, 
which is always a space-time scalar. Thus, the only boundary-bulk
propagators that we need are scalar and for a dimension $\Delta$ operator
are given by
\begin{equation}
K_\Delta (\vec{x} ,z) =C_\Delta  \bigg ( { z_0 \over z_0^2 +
(\vec{z}-\vec{x})^2 }\bigg )^\Delta
\qquad \quad
C_\Delta = {\Gamma (\Delta) \over \pi ^{d/2} \Gamma (\Delta - d/2)}, 
\end{equation}
for $\Delta > d/2$ and $C_{d/2} = \Gamma(d/2)/2\pi^{d/2}$.
Divergences in \AdS\ integrals can arise only when one or several 
integration points approach the boundary. As a bulk point $z$ approaches
the boundary, we have the following behaviors
\begin{eqnarray}
\label{asympt}
K_\Delta (\vec{x} ,z) 
   & \to & C_\Delta z_0 ^\Delta  { 1 \over (\vec{z}-\vec{x})^{2\Delta}} 
\qquad z_0 \to 0, \ \vec{z} \not= \vec{x}
\\
K_\Delta (\vec{x} ,z) 
   & \to &  z_0^{4-\Delta} \delta (\vec{z} - \vec{x})
\ \ \ \qquad \vec{z} \to \vec{x}
\end{eqnarray}
Bulk-to-bulk propagators, which occur in exchange graphs can have any spin
occurring in the table of descendents. For simplicity, we concentrate on 
the scalar bulk-to-bulk propagator $G_\delta (z,w)$ \cite{D'Hoker:1999mz}. 
We shall make use here only of its asymptotic behavior as one of the bulk
points approaches the boundary, and we have 
\begin{equation}
\label{asympt2}
G_\Delta (z,w) \to {2 w_0 ^\Delta \over (2 \Delta -d)(2\Delta -d-1)} \
K_\Delta (\vec{w}, z) \qquad {\rm as} \qquad w_0 \to 0
\end{equation}
Generally, supergravity vertices may involve derivative couplings. It
was shown in \cite{Freedman:1999tz} that the action of derivatives on
propagators may always be converted into non-derivative couplings up to a
multiplicative factor, which may vanish, but which will never diverge.

\subsection{Contact Graphs}

The convergence properties of \AdS\ integrals associated with contact
interactions and non-derivative coupling is simple. We consider a
correlator for operators of dimensions $\Delta$, $\Delta _i$,
$i=1,\dots,n\geq 2$, where each of the dimensions obeys the \AdS\
unitarity bound $\Delta \geq d/2$, and we assume that $\Delta \geq \Delta
_i$ for all $i$. We now regularize the \AdS\ integral by keeping $\Delta
_i$ set at their BPS values, while allowing $\Delta$ to be a general
complex number in the neighborhood of its BPS value.  The contact \AdS\
integral for a correlator with non-coincident points $x, \ x_i$, is given
by
\begin{equation}
I(\Delta, \Delta_i) 
=\int {d^{d+1}z \over z_0^{d+1}} K_\Delta (x,z)
\prod _{i=1} ^n K_{\Delta _i} (x_i,z) \, .
\end{equation}
By the unitarity bound on the $\Delta$'s, the integral is convergent in 
the region $z\to \infty$, and in view of (\ref{asympt}), it is convergent
as well when $z$ tends to any boundary point different from $x$ and
$x_i$.  As $z\to x_j$, we have
\begin{equation}
I(\Delta, \Delta_i) \sim {C_{\Delta } C_{\Delta _j} 
\over (x-x_j)^{2 \Delta}} 
\prod _{i\not=j} ^n {C_{\Delta _i} \over (x_i-x_j)^{\Delta _i}} 
\times 
\int {d^{d+1}z \over z_0^{d+1}} {z_0 ^{\Delta + \Delta _1 + \cdots + 
\Delta _n}
\over (z_0^2 + (z-x_j)^2)^{\Delta _j}}
\end{equation}
This integral is convergent around $z\to x_j$ as long as $\Delta + 
\Delta _1 + \cdots + \Delta _n - 2 \Delta _j >0$, which is guaranteed here
by the fact that
$\Delta \geq \Delta _i$ for all $i=1,\dots ,n$. 
As $z\to x$ on the other hand, we have
\begin{equation}
\label{extremalint}
I(\Delta, \Delta_i) \sim C_{\Delta }  
\prod _{i=1} ^n {C_{\Delta _i} \over (x_i-x)^{2\Delta _i}} 
\times 
\int {d^{d+1}z \over z_0^{d+1}} {z_0 ^{\Delta + \Delta _1 + \cdots + 
\Delta _n}
\over (z_0^2 + (z-x)^2)^{\Delta } }
\end{equation}
This integral is convergent as $z\to x$ as long as $\Delta < \Delta _1 + 
\cdots + \Delta _n$. Now, the group theory of the R-symmetry group (which
coincides with the isometry group of the sphere in AdS$\times$S)
guarantees that any correlator with $\Delta > \Delta _1 + \cdots + \Delta
_n$ vanishes identically. Thus, there is only one relevant divergence of
(\ref{extremalint}) which occurs precisely at $\Delta = \Delta _1 +
\cdots + \Delta _n$. The corresponding pole may be extracted exactly, and
we have
\begin{equation}
{\rm pole} \ I(\Delta, \Delta_i) = {1 \over \Delta_1 + 
\cdots + \Delta _n - \Delta} \times
\prod _{i=1} ^n { C_{\Delta _i} \over (x-x_i)^{2 \Delta _i}}
\end{equation}
as represented on Fig. 1.
Notice that the divergence structure and even the value of the pole 
divergence (up to the normalizations $C_{\Delta _i}$) are completely
independent of the
\AdS-space dimension $d$.
\EPSFIGURE{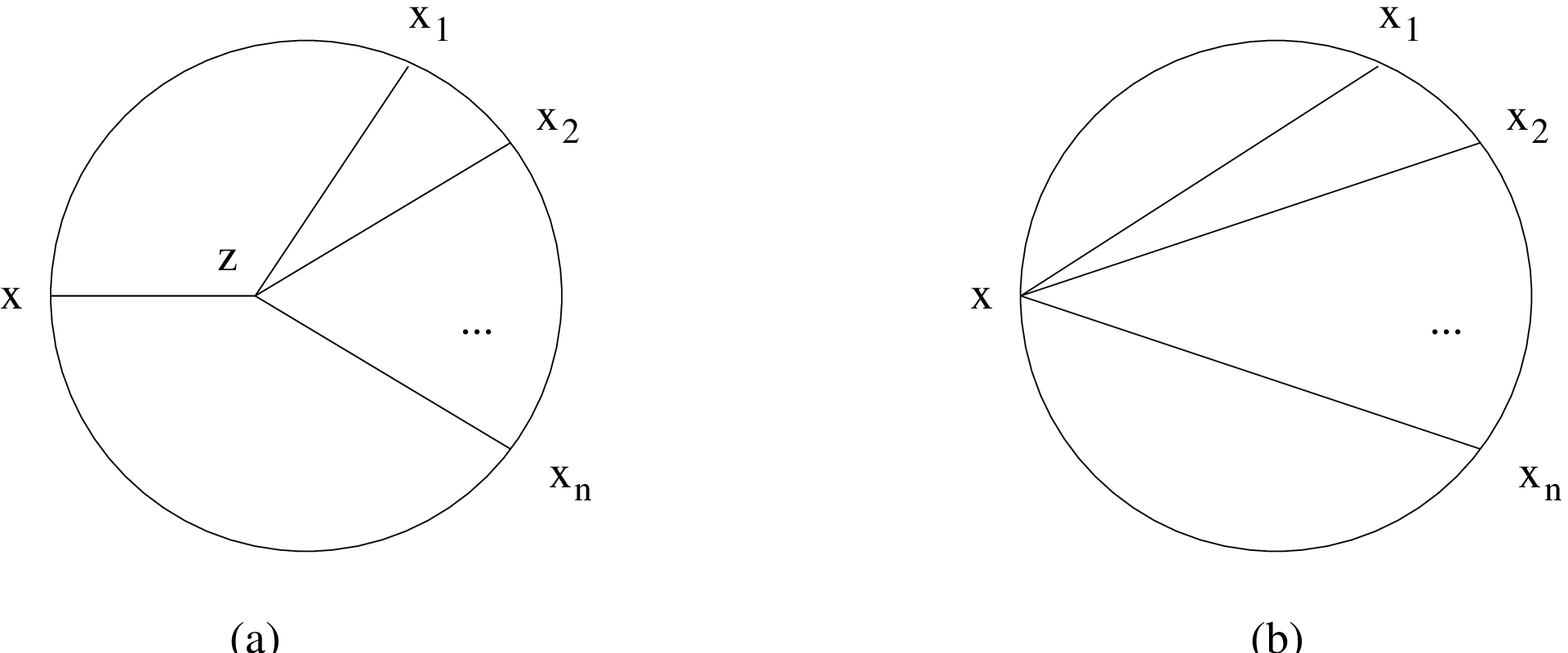,height=5cm}{Contact graph (a); Factorized
residue at $\Delta=\Delta_1+\dots+\Delta_n$~(b).}
 
\subsection{Exchange Graphs}

The analysis of the convergence structure for exchange diagrams is 
completely analogous to that of the contact terms. Remarkably, the
structure is again independent of the \AdS-space dimension
$d$, and we may carry over to general $d$ the criteria of divergence
given in \cite{D'Hoker:2000dm}. For an exchange graph with boundary points
$x_i$, $i=1,\dots ,n$ and superconformal primary (scalar) operators of
dimensions $\Delta _i$ at these points, we thus have the following criteria.

A simple pole divergence occurs in the z-integration over $\AdS_{d+1}$ if 
and only if the following two conditions are satisfied:

\begin{itemize}

\item Either $z$ approaches a point $x_i$ that is connected to $z$ by a
boundary-to-bulk propagator or it approaches a point $x_i$ that is 
connected to $z$ by a string of propagators and bulk interaction points
$z_a$ and all points $z_a$ also approach $x_i$;

\item The vertex at $z$ is extremal and the highest dimension of the fields
entering the vertex is the one of the field that connects $z$ (directly or
through a string of propagators) with $x_i$.

\end{itemize}

The residue of the poles may be calculated in a recursive way. We begin
with the $z$-integration over the bulk vertex that is connected to the
external operator of the largest dimension $\Delta$. To have a pole, this
vertex must be extremal, and $\Delta$ must be larger than any of the
dimensions of propagators emanating from the $z$-vertex, including
bulk-to-bulk propagators. The corresponding exchange amplitude may then
be represented by
\begin{eqnarray}
E(\Delta, \delta _a, \Delta_i)  =\int {d^{d+1}z \over z_0^{d+1}}
K_\Delta (x,z)
\prod _{i=1} ^p K_{\Delta _i} (x_i,z) \prod _{a=1} ^q
\int {d^{d+1}z_a \over (z_a)_0^{d+1}} G_{\delta _a} (z,z_a) 
D_a (z_a,\{x_j\}_a)
\ \qquad
\end{eqnarray} 
where $D_a(z_a,\{x_j\}_a)$ is a reduced amplitude with $n_a+1$
external legs, graphically represented in
Fig. 2,  and $p+\sum_{a=1}^q n_a=n$. 
As the $z$-vertex is assumed to be extremal, we have $\Delta =
\Delta _1 + \cdots + \Delta _p+ \delta_{1} + \cdots +\delta _{q}$, 
and we shall allow $\Delta$ to relax
away slightly from this value so as to suitably analytically the AdS
integrals. The only divergence of the $z$-integral arises when $z\to x$,
which we analyze in parallel to the case of the contact graph. Using the
asymptotics of (\ref{asympt}) and (\ref{asympt2}), we isolate the only
divergence of this integral, which occurs when $\Delta = 
\Delta _1 + \cdots + \Delta _p+ \delta _1 +
\cdots + \delta _q$. The pole part is
\begin{eqnarray}
\label{extremalint3}
{\rm pole} E(\Delta, \delta _a, \Delta_i) & = & {1 \over \Delta -
\sum _a \delta _a - \sum _i \Delta _i } \times 
\prod _{i=1}^p {C_{\Delta _i} \over (x_i-x)^{2\Delta _i}} \\
&& \times
\prod _{a=1}^q {2 \over (2\delta _a -d) (2 \delta _a -d -1)}
\int {d^{d+1}z_a \over (z_a)_0^{d+1}} K_{\delta _a} (x,z_a) D_a
(z_a,\{x_j\}_a)\, .
\nonumber
\end{eqnarray}
This result is graphically represented in Fig. 2. Each factor of the
residue may now be treated iteratively using the same formulas. With the
help of it, we shall analyze the factorization properties of near-extremal
correlators in the next sections.
\EPSFIGURE{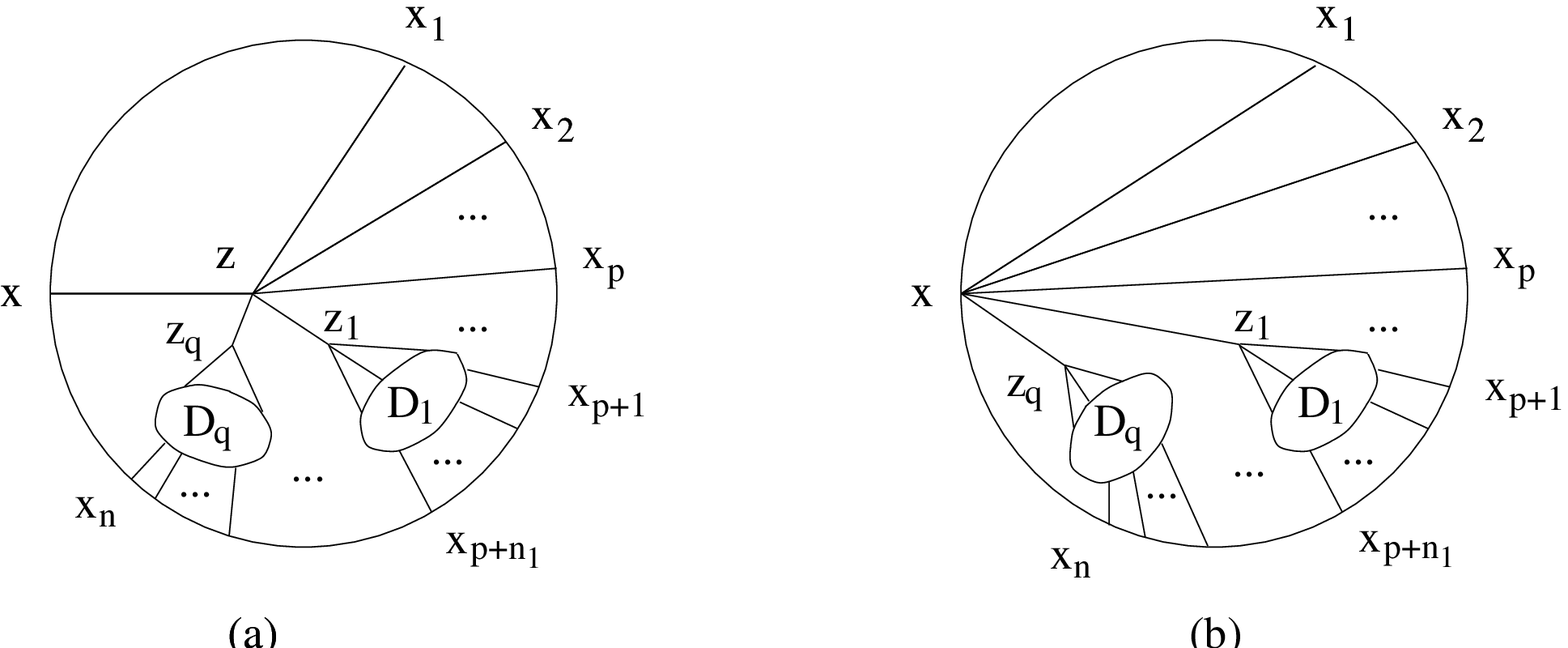,height=5cm}{Exchange graph (a); Factorized
residue at $\Delta=\Delta_1+\dots+\Delta_p+\delta_1+\dots+\delta_q$~(b).}

\section{Extremal Correlators}

Extremal correlators of 1/2 BPS operators $\O_{\Delta }$ of dimension 
$\Delta$ are of the form 
\begin{equation}
\label{extcorr}
\langle \O_{\Delta} (x) \O_{\Delta _1}
(x_1) \cdots \O_{\Delta _n} (x_n) \rangle 
\qquad \quad \Delta = \Delta _1 + \cdots + \Delta _n\, .
\end{equation}
We shall present arguments below that extremal correlation functions factorize
into a product of 2-point functions as follows
\begin{equation}
\label{fact}
\langle \O_{\Delta} (x) \O_{\Delta _1}
(x_1) \cdots \O_{\Delta _n} (x_n) \rangle 
= A(\Delta, \Delta _i;N) \prod _{i=1} ^n {1 \over (x-x_i)^{2 \Delta _i}}
\end{equation}
where the overall correlator strength $A$ may be expressed solely in terms 
of 2-point functions of 1/2 BPS operators. The first part of the argument
is based on the \AdS\ structure, and leads to the factorized form of the
correlator, while the second is based on the OPE and allows us to relate
the overall strength of the correlator to 2- and 3-point functions of 1/2
primary operators. From the AdS arguments, we also obtain the conjectures
on the extremal supergravity couplings, as given in (\ref{couplings2}) of
section~1.

\subsection{AdS Argument}

The fundamental assumption in the AdS argument is that M-theory on the 
\B47 and \C74 backgrounds is a finite theory at tree level, and that a
finite Maldacena dual conformal field theory at the boundary of $\AdS$
exists. The argument then proceeds inductively on the number
$n+1$ of operators entering the correlator (\ref{extcorr}).

For $n=2$, the extremal three point function must factorize by
conformal  invariance alone,
\begin{equation}
\label{three}
\langle \O_{\Delta } (x) \O_{\Delta _1}
(x_1) \O_{\Delta _2} (x_2) \rangle = A(\Delta , \Delta_1, \Delta _2;N) 
\prod _{i=1}^2 {1 \over (x-x_i)^{2\Delta _i}}\, .
\end{equation}
We also know from explicit supergravity calculations on \B47 in and 
\C74 in \cite{Corrado:1999pi,
Bastianelli:2000vm} that the extremal supergravity couplings vanish. The
product of the vanishing extremal supergravity coupling with the
divergent AdS integral accounts for the finite extremal correlator of
(\ref{three}). As shown in \cite{D'Hoker:1999ea}, one may turn this
argument around : the divergence of the extremal AdS integral together
with the finiteness assumption of M-theory on \B47 and \C74 imply that
the extremal supergravity coupling $\G(\Delta _1 + \Delta _2 ,
\Delta _1, \Delta _2)=0$.

For $n=3$, the extremal 4-point function, we have two types of \AdS\ 
graphs : a contact graph and exchange graphs. Purely from $SO(8)$ and
$USp(4)$ R-symmetry group theory for the \B47 and \C74 backgrounds, we
know that all bulk interaction vertices in each of these graphs have to
be extremal themselves. The two extremal 3-point couplings in the
exchange graphs produce two zeros which are multiplied by doubly
divergent AdS integrals. The result is finite and the pole contributions
from each AdS integral localize each bulk vertex  at the point $x$ on the
boundary. The result from the exchange graphs is a finite contribution of
the factorized form (\ref{fact}). There remains an extremal contact
graph, whose AdS integral diverges. Assuming finiteness again, this
implies that the extremal supergravity 4-point coupling must vanish, as
indeed conjectured in (\ref{couplings2}).

Next, assuming that all extremal $p$-point supergravity couplings vanish 
for $p\leq n$, it is easy to show that they must vanish for all $p \leq
n+1$. All bulk interaction vertices for exchange graphs are at most
$n$-point couplings and must be themselves extremal, and thus vanish by
the assumption of induction. Picking up the poles of the divergent AdS
integrals, one finds again a finite factorized form as in (\ref{fact}).
The remaining contact graph involves an $n+1$-point extremal supergravity
coupling, which is not yet known to vanish. However, the divergence of
the AdS integral together with the finiteness assumption again forces
also this extremal supergravity coupling to vanish, proving the induction
statement at order $n+1$.

Thus, there only remain the exchange graphs, in which every bulk vertex
is now extremal. Starting with the highest dimension boundary-to-bulk
propagator, we apply the recursive pole expressions of
(\ref{extremalint3}) and cancel the vanishing extremal coupling by the
pole of the AdS integral. The corresponding bulk vertex collapses onto
the boundary and the process can now be repeated for each of the factors.
There results, at the end of this recursive process the completely
factorized form of (\ref{extcorr}), with an overall factor $A$ which is
independent of $x_i$.

\subsection{Operator Product Expansion Argument}

In this subsection we show that, assuming the factorized space-time form 
of the extremal correlators as in (\ref{fact}), the overall normalization
$A(\Delta , \Delta _i;N)$ is related to 2- and 3-point functions of 1/2
BPS operators.

Recall that the ``single color trace" scalar operators $\O_k (x)$ have
dimension $\Delta = \K k$, where $\K = 1/2, 1, 2$ for \B47, \A5 and
\C74 respectively, while their R-symmetry Dynkin labels are
$(k,0,0,0)$ of $SO(8)$ for \B47, $(0,k,0)$ of $SU(4)$ for \A5 and $(0,k)$
of $USp(4)$, for \C74. They couple directly to the supergravity and
Kaluza-Klein modes with the corresponding quantum numbers. By taking the
OPE of such operators with suitable projections applied, we obtain
further ``multiple color trace" scalar operators with the same quantum
numbers. There exists an independent operator for every
partition of $\Delta $ into a sum of dimensions $\Delta _i$ of single
color trace operators obeying the unitarity bound $\Delta _i \geq d/2$.
We shall denote such partitions $\pi (\Delta)$ or simply by $\pi$ when no
confusion is possible. The corresponding operators are defined by
\begin{eqnarray}
\O_{\pi (\Delta)} (x) &=& \big [ \O_{\Delta _1}(x) \cdots \O_{\Delta _p} 
(x) \big ] \big | _{{\rm proj} \ 1/2 \ BPS} \\
\pi (\Delta) : \Delta &=& \Delta _1 + \cdots + \Delta _p\, .
\nonumber
\end{eqnarray} 
It is customary to normalize the single trace operators
$\O_\Delta (x)$ by their  2-point functions
\begin{equation}
\langle \O_\Delta (x) \O _{\Delta '} (y) \rangle = {\delta _{\Delta, 
\Delta '} \over (x-y) ^{2 \Delta}}\, .
\end{equation}
The normalization of the operators $\O_{\pi (\Delta)}$ is then determined
by the dynamics of the theory and  given by
\begin{equation}
\langle \O_{\pi (\Delta) (x)} \O _{\sigma (\Delta ')} (y) \rangle =
{\delta _{\Delta, \Delta '} M_{\pi,\sigma}
\over (x-y) ^{2 \Delta}}\, .
\end{equation}
For given $\Delta$, the matrix $M$ of all partitions is positive definite,
symmetric and depends only upon $\Delta $ and the number of colors $N$.

First, assuming the factorized form for the extremal correlators of
operators $\O_\Delta$, as in (\ref{fact}), the coefficient $A$ is given
by 
\begin{equation}
A(\Delta, \Delta _i;N) = M_{\pi, \sigma}
\qquad 
\pi (\Delta ) = \Delta, \quad \sigma (\Delta ) = \Delta _1 + \cdots + 
\Delta _n
\end{equation}
This follows directly by letting all points $x_i \to y$ and then using
the definition of $\O_{\pi (\Delta)}$ and the normalization of these 
operators.

Second, the argument may be generalized to the case of a correlation 
function of multi trace operators, characterized by non-trivial
partitions $\pi (\Delta)$. We assume that the factorized form holds for
single trace operators $\O_\Delta$.  This is enough to show that it holds
also for multi-trace operators. Suppose we wish to evaluate the correlator
\begin{equation}
\langle \O_{\pi (\Delta)} (x) \O_{\pi _1 (\Delta _1)} (x_1) \cdots \O_{\pi _n
(\Delta _n)}(x_n) \rangle
\qquad \quad\ ,
\Delta = \Delta _1 + \cdots + \Delta _n \, ,
\end{equation}
for given partitions 
\begin{equation}
\pi (\Delta ) = \delta ^{(1)} + \cdots + \delta ^{(p)}
\qquad \quad
\pi _i (\Delta _i) = \delta ^{(1)} _i + \cdots + \delta ^{(p_i)} _i \, .
\end{equation}
Clearly, this correlation function can be obtained from the correlator 
involving $\O_{\pi(\Delta)}$ and products of single trace operators only,
\begin{equation}
\langle \O_{\pi (\Delta)} (x) \ \prod _{i=1} ^n \ \prod _{a=1} ^{p_i} \
\O_{\delta ^{(a)} _i} (x_{i,a})  \rangle
\end{equation}
by letting $x_{i,a} \to x_i$ for all $a=1,\dots , p_i$, with unit 
coefficient of proportionality between the two correlators.

Third, it remains to link the operator of maximal dimension $\O_{\pi
(\Delta)}$  to single trace operators, in such a way that the correlator
may be evaluated from extremal correlators of single trace operators
alone. One may be tempted to view $\O_{\pi (\Delta)}(x)$ as the composite
of
$\O_{\delta ^{(1)}}(x) \cdots \O_{\delta ^{(1)}}(x)$, but this would give
rise to a correlation function which is not, in general, extremal. 
Instead, one must reconstruct the operator $\O_{\pi (\Delta)}$ from the 
most singular OPE term in the operator product expansion of operators of
a string of single trace operators, as follows
\begin{equation}
\O_{\Delta +\delta } (x) \O_{\tau (\delta)} (y)
=
{ 1 \over (x-y)^{2\delta} } \sum _{\pi} \Lambda _{\tau, \pi} \O_{\pi 
(\Delta)} (x) + {\rm less \ singular \ terms}
\end{equation}
where $\Lambda _{\tau, \pi}$ is a matrix that depends upon $\Delta$, 
$\delta$ and $N$. It may be calculated from the extremal 3-point function
and the matrix $M$ of 2-point functions. The dimension $\delta$ is taken
sufficiently large so that the relation may be inverted.

\section{General Near-Extremal Correlators}

We now discuss the case of near-extremal correlators of the form
\begin{equation}
\label{nextcorr}
\langle \O_{\Delta} (x) \O_{\Delta _1}
(x_1) \cdots \O_{\Delta _n} (x_n) \rangle 
\ ,\qquad \Delta + 2m {\cal K}= \Delta _1 + \cdots + \Delta _n\ , 
\end{equation}
where $m$ is an integer with $0 \leq m \leq n-2$, and ${\cal K}$ is the
unit  of dimension for superconformal primaries,
namely $\K=1/2$ for \B47, $\K=1$ for \A5 and $\K=2$ for \C74. 
The integrality of $m$ follows from the conservation 
of the $\Zint_2$ subgroup in the center of
the R-symmetry groups $SO(8)$, $SO(6)$ or $USp(4)$.
We shall generically denote such correlators by $E_{n+1}^m$.
For $m=0$, we recover the extremal correlation functions, which were
treated already in the preceding section. Next-to-extremal correlators
have $m=1$, and the bound above starts at 4-point correlators.
Next-to-next-to-extremal correlators necessitate 5 points or more, and so
on.

In the \A5 case, it was conjectured that such correlators $E_{n+1}^m$
decompose into a sum of products of non-renormalized 2- and 3-point
functions, and (for $m\geq 2$) renormalized higher-point functions. This
decomposition property has been checked on the gauge theory side up to
order $\g^2$  in \cite{D'Hoker:2000dm}. Precisely the same structure
emerges from the contribution of the exchange diagrams (for $n\geq 4$) on
the AdS side. The contact graph, which also arises on the AdS side, would
spoil this decomposition property, as it cannot be written as a sum of
factorized contributions in a non-trivial way. Assuming that the
decomposition property found at weak coupling extends to
strong coupling, we are naturally led to conjecture that the supergravity
couplings $\G(\Delta, \Delta_1, \dots, \Delta_{n})$ should vanish. This
has been shown by direct supergravity calculation for next-to-extremal
4-point couplings in \cite{Arutyunov:1999fb}.

Clearly, the arguments in favor of the vanishing of the sub-extremal
supergravity couplings ($m\geq 1$) are not quite as compelling as the
arguments given in favor of the vanishing of the extremal correlators,
where finiteness criteria played a deciding role. Nonetheless,
vanishing in the sub-extremal case appears to hold true and produces a
compelling global picture for structure of near-extremal correlators. 
In the \B47 and \C74 cases of interest to us, we do not have a {\it weak} 
coupling description at our disposal. What we do have is the fully
interacting strong coupling superconformal $\N=8$ and $(0,2)$ field
theories on the one side and the zero couplling (free) theories on the
other. But there is no family of superconformal (or even conformal) field
theories continuously connecting the two.  
By analogy with the $\N=4$ case of \cite{D'Hoker:2000dm}, it is clear that
the free $\N=8$ or $(0,2)$ theories yield the proposed decomposition of
the correlators. Since the free theory is disconnected from the fully
interacting theory of interest though, we cannot draw much support for
the proposed decomposition of the correlators from it. 

The situation is
however more promising on the AdS side: in complete analogy with
\cite{D'Hoker:2000dm},  we can show that the exchange diagrams
decompose into a sum of products of lower order correlation functions,
assuming that near-extremal supergravity couplings vanish. The
decomposition may be schematically represented as 
\begin{equation}
E_n^m\vert_{\rm exchange}=
\sum_{\{n_j,m_j\}} \prod_{i=1}^{n-m-1} E_{n_i}^{m_i}
\qquad {\rm with} \qquad
\sum_{i=1}^{n-m-1}n_i=2(n-1)-m
\end{equation}
with $\sum_{i=1}^{n-m-1}=m$.
The restriction $m\leq n-3$ ensures that each exchange
diagram decomposes into a sum of terms, each of which has at least two
factors, so that $n_i<n-1$. The arguments are completely parallel to
those given in \cite{D'Hoker:2000dm}, so we shall limit ourselves here to
presenting the cases of next-to-extremal correlators $E_{n+1}^1$.

We consider the next-to-extremal $n+1$-point functions
\begin{equation}
\langle \O_{\Delta} (x)
\O_{\Delta_1}(x_1) \cdots  \O_{\Delta_n} (x_n) \rangle
\qquad {\rm with} \qquad
\Delta =\Delta_1 + \cdots + \Delta _n + 2\K\, .
\end{equation}
We begin with $n=3$ and assume that the states propagating in one of the
exchange diagrams is a superconformal primary with dimension
$\Delta_e$. Group theory requires  
\be 
\Delta\leq \Delta_e+\Delta_1 \leq (\Delta_2+\Delta_3)+\Delta_1=
\Delta+2m\K
\end{equation}
with equality for extremal couplings. One of the two
vertices has to be extremal and hence have vanishing coupling. 
If the vertex $(\Delta\Delta_1\Delta_e)$ is extremal, the vanishing
coupling cancels the pole as the vertex approaches the boundary, leaving
a factorized result $\langle\O_{\Delta_1}\O_{\Delta_1}\rangle 
\langle\O_{\Delta_2}\O_{\Delta_3}\O_{\Delta_2+\Delta_3+2\K}\rangle$.
If it is the other vertex, the integral is finite and the net result
is zero. We thus have $E_4^1=E_2^0 E_3^1$.  Exchange of descendants may
be treated as in \cite{D'Hoker:2000dm}, and do not modify the picture. The
case $E_n^1$ is very similar: the only contribution comes when the vertex
linking the operator of highest dimension
$\O_{\Delta}$ to the tree is extremal and approaches the boundary.
The diagram then splits into two factors, one extremal and the other 
next-to-extremal. The latter further decomposes until one is left with
one 3 point function and $n-2$ two-point functions, the space-time
dependence of both of these being fixed by conformal symmetry alone.

The AdS integral for the extremal contact graph is divergent and
finiteness of tree-level supergravity thus forces the extremal
supergravity coupling to vanish. For next-to-extremal correlators,
the contact graph is finite, and merely spoils the factorizability of the
correlator. In analogy with the \A5 case, it is likely that the
factorization of near-extremal correlators is a consequence of
supersymmetry, and hence that the contact term should vanish on those
grounds. On the AdS side, it is remarkable that all exchange
diagrams exhibit the factorizability property, while the single contact
graph would not.

As was shown in \cite{D'Hoker:2000dm}, the same
reasoning applies to general near-extremal correlators. Again, their
exchange graphs factorize provided the contact graphs at lower orders are
absent. We are thus led to conjecture the vanishing of  near-extremal
supergravity couplings
\begin{equation}
\G(\Delta, \Delta_1, \dots, \Delta_{n})=0
\ ,\qquad \Delta + 2m {\cal K}= \Delta _1 + \cdots + \Delta _n
\end{equation}
for $m\leq n-2$.
It would be interesting to verify that statement at the level of 
4-point functions by adapting the analysis of \cite{Arutyunov:1999fb} to 
the  \B47 or \C74 case.

Assuming the above conjecture, the space-time form of the next-to-extremal
correlators may be written down exactly, up to a number of space-time
independent couplings $A_{ij} ^{(n)} (\Delta, \Delta _1,\dots , \Delta
_n;N)$ 
\begin{equation}
\langle \O_{\Delta} (x)
\O_{\Delta_1}(x_1) \cdots  \O_{\Delta_n} (x_n) \rangle
=
\sum _{i<j}^n A_{ij} ^{(n)}{(x-x_i)^{2 \K}
(x-x_j)^{2 \K} \over (x_i-x_j)^{2\K}} 
\prod _{k=1} ^n {1 \over (x-x_k)^{2 \Delta _k}} 
\end{equation}
Using the OPE of two of the operators, one may relate the couplings
$A_{ij} ^{(n)} (\Delta, \Delta _1,\dots,$ $\Delta _n;N)$ to the 2- and
3-point couplings of single- and multi-trace 1/2 BPS operators.

\acknowledgments

The authors acknowledge valuable discussions 
with J. Erdmenger, S. Ferrara, D. Freedman, M. Grisaru, 
S. Minwalla and M. Perez-Victoria. 
B. P. is grateful to UCLA Department of Physics for its warm hospitality
during part of this work.
The research of E. D. is supported in part by
NSF grant PHY-98-19686, and that of B. P. by 
the David and Lucile Packard Foundation
and the European TMR network ERBFMRX-CT96-0045.

\end{document}